\documentclass[10pt,twocolumn]{article}
\usepackage{graphicx}
\usepackage{amssymb}
\usepackage{amsmath}
\usepackage{bm}
\usepackage{epsf}
\usepackage{cancel}
\usepackage{tikz}

\begin{document}

\title{Probing Quantum Correlation Functions Through Energy Absorption Interferometry}

\author{S. Withington, C. N. Thomas, and D. J. Goldie \\ \\
Cavendish Laboratory, University Cambridge, J. J. Thomson Avenue, Cambridge}

\date{\today}

\maketitle

{\bf Abstract:} An interferometric technique is proposed for determining the spatial forms of the individual degrees of freedom through which a many body system can absorb energy from its environment. The method separates out the coherent excitations present at any given frequency; it is not necessary to infer modal content from spectra. The system under test is excited with two external sources, which create generalized forces, and the fringe in the total power dissipated is measured as the relative phase between the sources is varied. If the complex fringe visibility is measured for different pairs of source locations, the anti-Hermitian part of the complex-valued non-local correlation tensor can be determined, which can then be decomposed to give the natural dynamical modes of the system and their relative responsivities. If each source in the interferometer creates a different kind of force, the spatial forms of the individual excitations that are responsible for cross-correlated response can be found. The technique is a generalization of holography because it measures the state of coherence to which the system is maximally sensitive. It can be applied across a wide range of wavelengths, in a variety of ways, to homogeneous media, thin films, patterned structures, and to components such as sensors, detectors and energy harvesting absorbers.

\section{Introduction}

Quantum correlation functions \cite{ref1}, and their related Green's functions, play a central role in solid-state physics. They describe dynamical behaviour, and reveal internal order, whilst preserving the exchange symmetries of the constituent particles. The Hermitian and anti-Hermitian parts of retarded correlation functions describe reactive and dissipative processes respectively, and the anti-Hermitian parts also characterise the fluctuations that are present when thermal systems are observed passively \cite{ref2a,ref2b}. Retarded Green's functions are central, through Landauer's formalism \cite{ref3}, to determining the transmissive channels available in multiport quantum networks.

In this paper we propose a technique, called Energy Absorption Interferometry (EAI), for measuring the anti-Hermitian parts of retarded correlation functions. Once this has been done it is possible to determine the spatial forms of the individual coherent excitations through which a many body system can absorb energy from its environment: individual plasma oscillations, current distributions, spin waves, phonon modes, etc. EAI also allows the spatial forms of the individual coherent excitations that connect generalized forces of different kinds to be determined.

The basic idea is to excite the system under test with two external sources, and then to measure the fringe in the total power dissipated as the relative phase between the sources is varied. If the complex fringe visibility is measured for different pairs of source locations, and where appropriate polarisations, the anti-Hermitian part of the complex-valued non-local correlation tensor can be determined, which can then be decomposed to give the natural dynamical modes of the system and their relative responsivities. The method separates out the coherent excitations present at any given single frequency; it is not necessary to infer modal content from spectra. Our proposed technique is essentially a generalization of holography because it measures the state of coherence to which the system under test is maximally sensitive.

There is a wide variety of reasons why it is important to know the allowed, collective excitations of many body systems. In the case of electromagnetic \cite{ref4a,ref4b,ref4c,ref4d,ref4e,ref4f}, elastic, piezo-electric, and acoustic sensors such as sonar \cite{ref5a,ref5b,ref5c,ref5d}, it is essential to know the number, efficiencies and precise forms of the individual modes through which the device can absorb energy. A scanned measurement with a single source can only determine the overall power reception pattern, it cannot determine the amplitude, phase and polarisation patterns of the individual modes that make up the total response.

In the case of microwave and optical photon-counting detectors for quantum communications \cite{ref6a,ref6b,ref6c,ref6d}, it is essential to avoid, or at least terminate carefully, electromagnetic modes that can only couple noise and stray light into the detector. In the case of energy harvesting components, antenna arrays and absorbers, including micro-mechanical devices \cite{ref7a}, it is essential to maximise the number of degrees of freedom available for collecting power. The same considerations apply to near-field energy and information transfer between separated or overlapping volumes \cite{ref8a,ref8b,ref8c}.
In the case of qubits for quantum computing \cite{ref9a,ref9b}, which may be based on electromagnetic, spin \cite{ref10a,ref10b}, or mechanical \cite{ref11a,ref11b} resonators, it is essential to understand the number, nature and origin of the mechanisms that couple the active elements to their passive environments, causing decoherence.

Because we describe EAI in terms of generalized conjugate variables, it has wide applicability. It can be implemented over a wide range of wavelengths, to homogeneous media, thin films, nano-patterned structures, classical and quantum metamaterials \cite{ref11c} , and to individual components and arrays. Unlike passive observations of thermal fluctuations, which through the fluctuation-dissipation theorem \cite{ref12a,ref12b,ref12c} also contain information about correlation functions \cite{ref13a,ref13b,ref13c,ref13d}, the method achieves high signal-to-noise ratios by driving the system under test with external sources. Low-power sources can be used to probe systems in near-equilibrium, and high-power sources can be used probe the differential behaviour of systems in highly non-equilibrium states. The method can also be used as a convenient tool for exploring and characterising the behaviour of numerical many-body simulations.

\section{Correlation functions}

If a generalized external classical force ${\bf F}^{n}( {\bf r}, t)$ acts on a many body quantum system, the change in the Hamiltonian is
\begin{equation}
\label{eqn1-1}
\hat{H}_{e} (t) = \int_{{\cal V}_{n}} d^{3} {\bf r} \, {\bf F}^{n}( {\bf r}, t) \cdot \hat{\bf F}^{n} ({\bf r}).
\end{equation}
Superscript $n$ denotes a specific generalized force within some set: electric scalar potential,  magnetic vector potential, magnetic field, elastic force, etc. Each generalized force is associated, through (\ref{eqn1-1}), with a quantum observable $\hat{\bf F}^{n} ({\bf r})$, that determines the forces contribution to the total energy. For reasons that will become clear, the domain of integration, ${\cal V}_{n}$, is indicated explicitly. The domains corresponding to different forces can be the same, overlapping or completely disjoint.

According to Kubo \cite{ref2a}, the expectation value of the change in $\hat{\bf F}^{m} ({\bf r},t)$ at space-time point $({\bf r},t)$ when generalized force ${\bf F}^{n} ({\bf r}',t')$ is applied is
\begin{equation}
\label{eqn1-2}
\langle \Delta  \hat{\bf F}^{m} ({\bf r},t) \rangle  = \int_{-\infty}^{+\infty} dt'   \int_{{\cal V}_{n}}  d^{3} {\bf r}' \, \, \overline{\overline{\bm \chi}}^{\, mn} ({\bf r},t; {\bf r}',t') \cdot {\bf F}^{n} ({\bf r}',t'),
\end{equation}
which is a generalized displacement. The spatial vector components $k,l$ of the generalized susceptibility tensor $\overline{\overline{\bm \chi}}^{mn} ({\bf r},t; {\bf r}',t')$ are given by the retarded correlation functions
\begin{equation}
\label{eqn1-3}
\chi_{kl}^{\, mn} ({\bf r},t; {\bf r}',t') = - \frac{i}{\hbar} \theta ( t - t') \langle \left[ \hat{F}^{mH}_{k} ({\bf r},t), \hat{F}^{n H}_{l} ({\bf r}', t')   \right] \rangle.
\end{equation}
We shall use dyadic notation, denoted by a double overline, for spatial vector operators, which does not preclude the possibility that one or both of the generalized forces may be scalars. $\hat{F}^{m H}$ and $\hat{F}^{n H}$ are operators in the Heisenberg picture,  $\langle \, \, \rangle$ denotes the expectation over the grand canonical ensemble using the effective Hamiltonian, which includes chemical potential, $\left[ \, \, \right]$ is the commutator, or anticommutator where appropriate, and the step function $\theta ( t - t')$ ensures causal response. ${\bf F}^{m} ({\bf r},t)$ and $\langle \Delta  \hat{\bf F}^{m} ({\bf r},t) \rangle$ form conjugate pairs, and so the formalism is general. Kubo's expression provides a way of calculating macroscopic response functions using quantum-statistical methods, and therefore it contains information about the spatial and temporal excitations allowed.

When a generalized force is applied, the expectation value of the instantaneous rate of work done is given by
\begin{equation}
\label{eqn1-4}
P(t)  = \int_{{\cal V}_{m}}  d^{3} {\bf r} \, {\bf F}^{m}( {\bf r}, t) \cdot \frac{d \langle \Delta \hat{\bf F}^{m} ({\bf r},t) \rangle}{dt},
\end{equation}
where $\langle \Delta \hat{\bf F}^{m} ({\bf r},t) \rangle$ is the expectation value of the resultant change in the operator on which ${\bf F}^{m}( {\bf r}, t)$ acts. Using (\ref{eqn1-2}) in (\ref{eqn1-4}), and allowing for two different kinds of force to be present simultaneously, $m,n \in 1,2$, the time averaged rate of work done becomes
\begin{align}
\label{eqn1-5}
\overline{P} &   =   \lim_{T \rightarrow \infty} \frac{1}{T} \int_{-T/2}^{T/2}  dt \int_{-\infty}^{+\infty}  dt'
\sum_{mn} \int_{{\cal V}_{m}} d^{3}{\bf r} \int_{{\cal V}_{n}} d^{3}{\bf r}'   \\ \nonumber
 & \hspace{15mm} {\bf F}^{m}( {\bf r}, t) \cdot \frac{ d \, \overline{\overline{\bm \chi}}^{\, mn} ({\bf r},t; {\bf r}',t') }{dt} \cdot {\bf F}^{n} ({\bf r}',t').
\end{align}
The diagonal terms $m = n$ give the powers dissipated by the forces individually, whereas the off-diagonal terms $m \neq n$ arise because the application of one force can result in a perturbation of the quantum observable associated with the other force. In some cases, the time averaging, over $T$, should be replaced by a convolution integral representing post-measurement filtering.

Often, we are only interested in probing self correlations $\overline{\overline{\bm \chi}}^{\, mm} ({\bf r},t; {\bf r}',t')$: for example electric or magnetic susceptibility. Sometimes, we are interested in measuring cross correlations, $\overline{\overline{\bm \chi}}^{\, mn} ({\bf r},t; {\bf r}',t')$:  for example, ferro-electric susceptibility. One can never measure all correlations associated with all possible physical variables, and so it will be necessary to extract subspaces corresponding to the quantities of interest.

Use the following spectral decomposition for the susceptibility tensor
\begin{align}
\label{eqn1-6}
\overline{\overline{\bm \chi}}^{\, mn} ({\bf r}, t; {\bf r}',t') & = \frac{1}{(2 \pi)^{2}}\int_{-\infty}^{+\infty} d \omega \int_{\infty}^{+\infty} d \omega' \\ \nonumber
& \overline{\overline{\bm \chi}}^{\, mn} ({\bf r},\omega; {\bf r}',\omega')  \exp \left[- i \omega t \right] \exp \left[ + i \omega' t' \right],
\end{align}
and require that the applied forces are time harmonic
\begin{align}
\label{eqn1-7}
 {\bf F}^{n}({\bf r},t)& = {\bf F}_{0}^{n}({\bf r}) \exp \left[ - i \omega_{0} t \right] + {\bf F}_{0}^{n \ast}({\bf r}) \exp \left[ + i \omega_{0} t \right].
\end{align}

For presentational simplicity, assume that the Hamiltonian of the unperturbed system is constant, so that the correlation function depends only on time differences, $\overline{\overline{\bm \chi}}^{\, mn} ({\bf r},{\bf r}',t-t') \equiv \overline{\overline{\bm \chi}}^{\, mn} ({\bf r},t;{\bf r}',t')$, and the spectral representation  becomes diagonal $ \overline{\overline{\bm \chi}}^{\, mn} ({\bf r},\omega; {\bf r}',\omega') = \overline{\overline{\bm \chi}}^{\, mn} ({\bf r},{\bf r}',\omega) 2 \pi \delta ( \omega - \omega')$. In this case, (\ref{eqn1-5}) becomes
\begin{align}
\label{eqn1-8}
\overline{P}  = & -i \omega_{0} \sum_{mn} \int_{{\cal V}_{n}} d^{3}{\bf r} \int_{{\cal V}_{n}} d^{3}{\bf r}'  \lim_{T \rightarrow \infty} \frac{1}{T} \int_{-T/2}^{T/2} dt \, \\ \nonumber
& \hspace{2mm} \Big\{ {\bf F}_{0}^{m}({\bf r}) \cdot \overline{\overline{\bm \chi}}^{\, mn} ({\bf r},{\bf r}',\omega_{0})  \cdot {\bf F}_{0}^{n}({\bf r}') \,   \exp \left[ - i  2 \omega_{0}  t \right]   \\ \nonumber
& - {\bf F}_{0}^{m \ast}({\bf r}) \cdot \overline{\overline{\bm \chi}}^{\, mn} ({\bf r},{\bf r}',- \omega_{0})  \cdot {\bf F}_{0}^{n \ast}({\bf r}')   \, \exp \left[ + i  2 \omega_{0} t \right]  \\ \nonumber
&  - {\bf F}_{0}^{m} ({\bf r}) \cdot \overline{\overline{\bm \chi}}^{\, mn} ({\bf r},{\bf r}',- \omega_{0})  \cdot {\bf F}_{0}^{n \ast}({\bf r}')  \\ \nonumber
&  + {\bf F}_{0}^{m \ast}({\bf r}) \cdot  \overline{\overline{\bm \chi}}^{\, mn} ({\bf r},{\bf r}',\omega_{0}) \cdot {\bf F}_{0}^{n}({\bf r}')  \Big\}.
\end{align}

For long integration times, $T \rightarrow \infty$, the first two terms disappear, and
\begin{align}
\label{eqn1-9}
\overline{P}   =  -i \omega_{0} & \sum_{mn} \int_{{\cal V}_{m}} d^{3}{\bf r} \int_{{\cal V}_{n}} d^{3}{\bf r}' \\ \nonumber
&  \hspace{2mm} \Big\{ {\bf F}_{0}^{m \ast}({\bf r}) \cdot   \overline{\overline{\bm \chi}}^{\, mn} ({\bf r},{\bf r}', \omega_{0}) \cdot {\bf F}_{0}^{n}({\bf r}')  \\ \nonumber
&  - {\bf F}_{0}^{m}({\bf r}) \cdot  \overline{\overline{\bm \chi}}^{\, mn \ast} ({\bf r},{\bf r}', \omega_{0}) \cdot {\bf F}_{0}^{n\ast}({\bf r}')  \Big\},
\end{align}
where we have used $\overline{\overline{\bm \chi}}^{\, mn} ({\bf r},{\bf r}',- \omega_{0}) = \overline{\overline{\bm \chi}}^{\, mn \ast} ({\bf r},{\bf r}', \omega_{0})$.

Each term is a scalar, and so the transpose can be taken without changing the result.  Taking the transpose of the second term, and swapping the dummy variables $m$ and $n$, and ${\bf r}$ and ${\bf r}'$, gives
\begin{align}
\label{eqn1-10}
\overline{P}   =  2 \omega_{0} & \sum_{mn} \int_{{\cal V}_{m}} d^{3}{\bf r} \int_{{\cal V}_{n}} d^{3}{\bf r}' \\ \nonumber
& {\bf F}_{0}^{m \ast}({\bf r}) \cdot  \overline{\overline{\bm D}}^{\, mn} ({\bf r},{\bf r}',\omega_{0}) \cdot {\bf F}_{0}^{n}({\bf r}'),
\end{align}
where
\begin{align}
\label{eqn1-11}
\overline{\overline{\bm D}}^{\, mn} ({\bf r},{\bf r}',\omega_{0}) & = \left[ \frac{   \overline{\overline{\bm \chi}}^{\, mn} ({\bf r},{\bf r}',\omega_{0}) -   \overline{\overline{\bm \chi}}^{\, nm \dagger} ({\bf r}',{\bf r},\omega_{0})}{2i} \right].
\end{align}
$\overline{\overline{\bm \chi}}^{\, nm \dagger} ({\bf r}',{\bf r},\omega_{0})$ is the adjoint of $\overline{\overline{\bm \chi}}^{\, mn} ({\bf r},{\bf r}',\omega_{0})$. $\overline{\overline{\bm D}}^{\, mn} ({\bf r},{\bf r}',\omega_{0})$ is the anti-Hermitian part of the susceptibility tensor, rendered Hermitian by the factor $i$ in the denominator. (\ref{eqn1-10}) is the average dissipated power when the spatial response is non-local, and the temporal response is stationary. It reduces to well-known expressions in the appropriate limits; for example spatial shift invariance.

Suppose that each applied classical force is itself a statistical quantity defined over an ensemble. (\ref{eqn1-10}) is a scalar, and so taking the trace on both sides, rotating ${\bf F}_{0}^{m \ast}({\bf r})$ to the right, and then calculating the classical average, $\langle \, \rangle$, gives
\begin{align}
\label{eqn1-12}
\langle \overline{P} \rangle   = & 2 \omega \sum_{mn} \int_{{\cal V}_{m}} d^{3}{\bf r} \int_{{\cal V}_{n}} d^{3}{\bf r}' \\ \nonumber
& \overline{\overline{\bm D}}^{\, mn} ({\bf r},{\bf r}',\omega) \cdot \cdot \, \overline{\overline{\bf F}}^{nm \dagger} ({\bf r}',{\bf r},\omega),
\end{align}
where double-dot notation is used to denote the contraction of the vectorial parts of the tensors. $\overline{\overline{\bf F}}^{\, nm \dagger} ({\bf r}',{\bf r},\omega) = \langle {\bf F}_{0}^{n}({\bf r}') {\bf F}_{0}^{m \ast}({\bf r}) \rangle$ is a tensor field that describes the spatial state of coherence of the applied generalized forces. Strictly, (\ref{eqn1-12}) is the average absorbed power when the applied forces are described in terms of slowly varying analytic signals \cite{ref14}. For broadband forces, (\ref{eqn1-12}) is a spectral power, and should be integrated over $\omega$.

(\ref{eqn1-12}) shows that the total dissipated power is given by the full contraction of two tensor fields to a scalar: one of which characterises the ability of the many body system to absorb energy, and the other characterises the spatial state of coherence of the applied forces. (\ref{eqn1-12}) is formally an inner product in a mixed tensor space, and so the measured power is given by the projection of a tensor that describes the state of coherence of the applied forces onto a tensor that describes the state of coherence to which the system is maximally receptive. This point will be discussed later.

\section{Absorption Interferometry}

\subsection{Self correlations}

\begin{figure}[h]
\noindent \begin{centering}
\includegraphics[trim = 2.5cm 18cm 11cm 4cm,width=60mm]{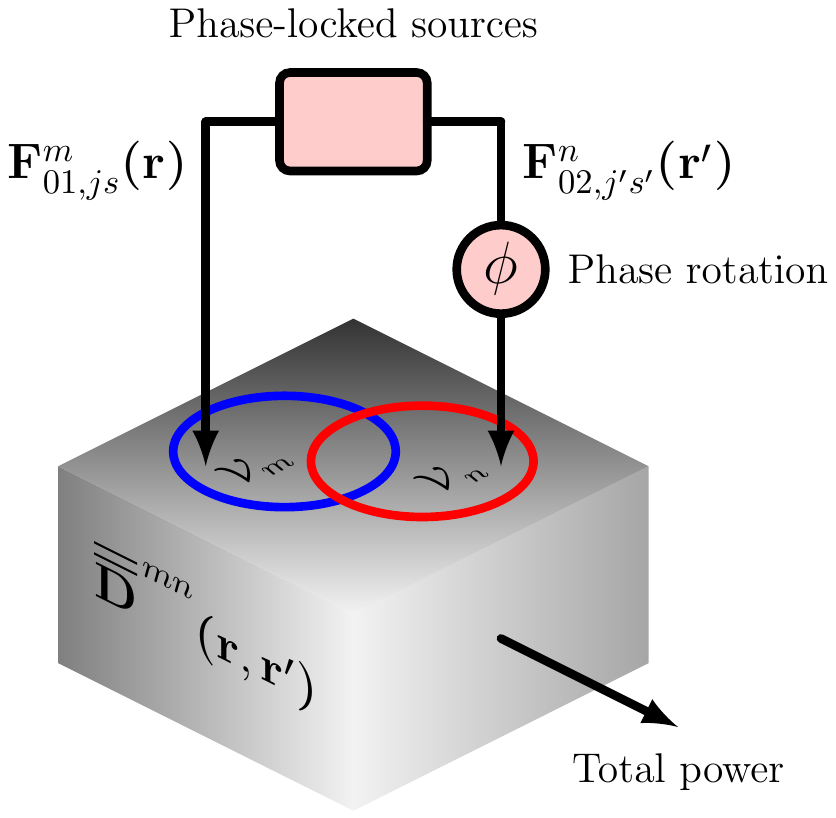}
\par\end{centering}
\caption{\small Energy Absorption Interferometer. Two phase-locked sources produce generalized forces ${\bf F}_{01,js}^{m} ({\bf r})$ and  ${\bf F}_{02,j's'}^{n} ({\bf r})$. The time-averaged total power displays a fringe as the differential phase $\phi$ is varied. The complex visibilities for different source locations enable the complex-valued system response tensor $\overline{\overline{\bf D}}^{\, mn} ({\bf r},{\bf r}')$ to be determined.}
\label{fig1}
\end{figure}

Consider the situation where two external, coherent, phase-locked sources of the same kind, $n=m=1$, are used to excite a system: Figure~\ref{fig1},
\begin{align}
\label{eqn2-1}
{\bf F}_{0}^{1} ({\bf r}) & =  {\bf F}_{01,js}^{1} ({\bf r}) + {\bf F}_{02,j's'}^{1} ({\bf r}) \exp \left[ i \phi \right] \\ \nonumber
{\bf F}_{0}^{2} ({\bf r}) & = {\bf 0}.
\end{align}
${\bf F}_{01,js}^{1} ({\bf r})$ is the vector force produce by the first source, denoted by the first subscript, when it is placed at sample position $j$ and in polarisation state $s$: for example, an electric or magnetic dipole. ${\bf F}_{02,j's'}^{1} ({\bf r})$ is the same quantity for the second source, but now its relative phase $\phi$ can be varied by the experimenter. In this case, ${\bf F}_{0}^{2} ({\bf r}) = {\bf 0}$ because there are no sources of the second kind.

Because the external sources are fully coherent and phased locked, the dissipated power is given by (\ref{eqn1-10}),
\begin{align}
\label{eqn2-2}
 \overline{P}_{js,j's'}  &  =   2 \omega_{0} \int_{{\cal V}_{1}} d^{3}{\bf r} \int_{{\cal V}_{1}} d^{3}{\bf r}' \\ \nonumber
 & \hspace{2mm} \Big\{ {\bf F}_{01,js}^{1 \ast} ({\bf r}) \cdot  \overline{\overline{\bm D}}^{\, 11} ({\bf r},{\bf r}',\omega_{0}) \cdot {\bf F}_{01,js}^{1} ({\bf r}') \\ \nonumber
 & + {\bf F}_{02,j's'}^{1 \ast} ({\bf r})  \cdot  \overline{\overline{\bm D}}^{\, 11} ({\bf r},{\bf r}',\omega_{0}) \cdot {\bf F}_{01,js}^{1} ({\bf r}') \exp \left[ - i \phi \right] \\ \nonumber
 & +  {\bf F}_{01,js}^{1 \ast} ({\bf r}) \cdot  \overline{\overline{\bm D}}^{\, 11} ({\bf r},{\bf r}',\omega_{0}) \cdot {\bf F}_{02,j's'}^{1} ({\bf r}') \exp \left[ i \phi \right]  \\ \nonumber
 & + {\bf F}_{02,j's'}^{1 \ast} ({\bf r}) \cdot  \overline{\overline{\bm D}}^{\, 11} ({\bf r},{\bf r}',\omega_{0}) \cdot {\bf F}_{02,j's'}^{1} ({\bf r}') \big\},
\end{align}
which follows because only one term is present in the sum over $m,n$. The domains of integration are now the same. $\overline{\overline{\bm D}}^{\, 11} ({\bf r},{\bf r}',\omega_{0})$ is Hermitian, and so the first and last terms are real, and independent of the phase difference between the sources. The first term is the total power absorbed from the source at position $j$ and in polarisation $s$, whereas the last term is the total power absorbed from the source at position $j'$ in polarisation $s'$. The second and third terms are complex scalars, and the complex conjugates of each other.

\begin{figure}[h]
\noindent \begin{centering}
\includegraphics[trim = 2.5cm 21cm 11cm 4cm,width=80mm]{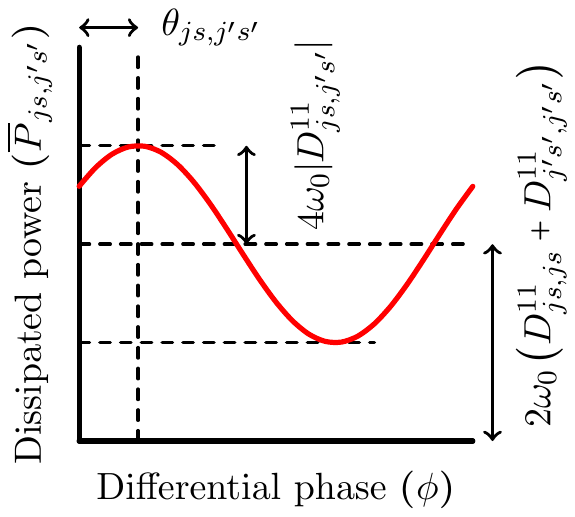}
\par\end{centering}
\caption{\small Fringe in the dissipated power $\overline{P}_{js,j's'}$ as relative phase between sources $\phi$ is varied. The average is given by the sum of the powers dissipated by the sources individually $2 \omega_{0}  \left( D^{11}_{js,js} + D^{11}_{j's',j's'}  \right)$, the height of the fringe is given by the magnitude of the cross correlation $4 \omega_{0} | D^{11}_{js,j's'}|$, and the phase of the fringe is given by the phase of the cross correlation $\theta_{js,j's'}$.}
\label{fig3}
\end{figure}

The dissipated power can be written
\begin{align}
\label{eqn2-3}
 \overline{P}_{js,j's'} = 2 \omega_{0} \left[ D^{11}_{js,js} \right. & + D^{11}_{j's',j's'} \\ \nonumber
 & \left. + 2 | D^{11}_{js,j's'} | \cos \left(  \phi + \theta_{js,j's'} \right) \right],
\end{align}
where $| D^{11}_{js,j's'} |$ and $\theta_{js,j's'}$ are the amplitudes and phases of
\begin{align}
\label{eqn2-4}
D^{11}_{js,j's'}   = & \int_{{\cal V}_{1}} d^{3}{\bf r} \int_{{\cal V}_{1}} d^{3}{\bf r}' \\ \nonumber
& {\bf F}_{01,js}^{1 \ast} ({\bf r}) \cdot  \overline{\overline{\bm D}}^{11} ({\bf r},{\bf r}',\omega_{0}) \cdot {\bf F}_{02,j's'}^{1} ({\bf r}'),
\end{align}
which are the matrix elements of the dissipative part of the susceptibility tensor in the vector space, strictly the dual space, of the sources. As the phase is varied, the dissipated power displays a fringe, Figure~\ref{fig3}, which gives the complex matrix elements. In practice, it is not necessary to sweep out each fringe explicitly, but it is sufficient to switch between two states $\phi = 0$ and $\phi = \pi/2$ to record the real and imaginary parts of the matrix element directly. In experimental work, we have found it convenient to run the sources at slightly different frequencies, and then to use a lock-in amplifier to measure the real and imaginary parts of the modulation directly \cite{ref15}.

If the fringe is recorded for enough source locations and polarisation states, to be quantified later, the complex-valued response tensor can be found in the vector space of the sources. Suppose that the sources are moved throughout some scanning region, volume or surface, leading to a total of $K$ sample positions and polarisations. The impressed forces then form a basis $\mathbb{F} = \{ {\bf F}_{0,k}({\bf r}), \forall \, k \in 1 \cdots K \}$, where different $k$ correspond to different combinations of $j$ and $s$. The resulting basis is general: it is not necessary to use the same polarisation states, or indeed orthogonal polarisation states, at the sample positions, which helps devise simple scanning strategies.

In cases where the sources produce point-like unidirectional forces, say mechanical probes,
\begin{align}
\label{eqn2-5}
{\bf F}_{01,js}^{1} ({\bf r}) & =  F_{01,s}^{1} \delta({\bf r} - {\bf r}_{j}) \hat{\bf x}_{s} \\ \nonumber
{\bf F}_{02,j's'}^{1} ({\bf r}) & = F_{02,s'}^{1} \delta({\bf r} - {\bf r}_{j'}) \hat{\bf x}_{s'}
\end{align}
can be substituted into (\ref{eqn2-4}) to yield
\begin{align}
\label{eqn2-6}
D^{11}_{js,j's'}  = F_{01,s}^{1 \ast}  D^{11}_{ss'} ({\bf r}_{j},{\bf r}_{j'}',\omega_{0}) F_{02,s'}^{1},
\end{align}
and the experiment measures directly the corresponding vector component of the spatial response tensor at the positions of the sources. The spatial coherence function can be traced out by moving the probes.

At the other extreme, where the sources provide spatially uniform forces in orthogonal directions, as in the case of the magnetic fields produced by orthogonal Helmholtz pairs,
\begin{align}
\label{eqn2-7}
{\bf F}_{01,s}^{1} ({\bf r}) & =  F_{01,s}^{1} \hat{\bf x}_{s} \\ \nonumber
{\bf F}_{02,s'}^{1} ({\bf r}) & = F_{02,s'}^{1} \hat{\bf x}_{s'}.
\end{align}
(\ref{eqn2-4}) then gives
\begin{align}
\label{eqn2-8}
D^{11}_{s,s'}  = F_{01,s}^{1 \ast} \Big\{ \int_{{\cal V}_{1}} d^{3}{\bf r} \int_{{\cal V}_{1}} d^{3}{\bf r}'
D^{11}_{ss'} ({\bf r}_{j},{\bf r}_{j'}',\omega_{0}) \Big\} F_{02,s'}^{1},
\end{align}
which shows that fringes are formed in the total dissipated power as the phase between the fields, currents in the orthogonal Helmholtz pairs, is varied. As will be seen later, this allows the directional forms of the individual degrees of freedom that make up the total, spatially integrated, directional response to be determined.

In general, the basis functions are neither orthogonal nor uniform over ${\cal V}_{1}$: for example the sampling fields produced by a scanned electric or magnetic dipole. The spatial susceptibility tensor must then be reconstructed through
\begin{equation}
\label{eqn2_9}
\overline{\overline{\bm D}}^{11} ({\bf r},{\bf r}',\omega_{0}) \approx \sum_{k,k'} D_{k,k'}^{11} \widetilde{\bf F}_{01,k}^{1}({\bf r}) \widetilde{\bf F}_{02,k'}^{1 \ast}({\bf r}')
\mbox{.}
\end{equation}
where $\widetilde{\bf F}_{01,k}^{1}({\bf r})$ is the dual of ${\bf F}_{01,js}^{1}({\bf r})$. The dual set $\widetilde{\mathbb{F}} = \{ \widetilde{\bf F}_{0,k}({\bf r}), \forall \, k \in 1 \cdots K \}$ can be found numerically, see later, once the functional forms of the impressed force are known. This scheme applies even if the two sources in the interferometer do not produce the same force distributions, say because they are not identical.

The source fields and their duals span the same vector space. (\ref{eqn2_9}) may, however, be an approximation because it is not generally known whether $\mathbb{F}$ is complete, over complete or under complete with respect to the degrees of freedom in $\overline{\overline{\bm D}}^{11} ({\bf r},{\bf r}', \omega_{0})$. Reconstruction using the dual functions covers all possibilities, giving the best orthogonal metric projection when the basis is under complete. The process of reconstructing $\overline{\overline{\bm D}}^{11} ({\bf r},{\bf r}', \omega_{0})$ using the dual set $\widetilde{\mathbb{F}}$ amounts to `deconvolving' the probe field patterns from the measurements.

\subsection{Cross correlations}

In some cases, the primary need is to determine the response tensor corresponding to two different kinds of generalized force. Interferometry is then carried out using two different kinds of source:
\begin{align}
\label{eqn2-10}
{\bf F}_{0}^{1} ({\bf r}) & =  {\bf F}_{01,js}^{1} ({\bf r}) \\ \nonumber
{\bf F}_{0}^{2} ({\bf r}) & = {\bf F}_{02,j's'}^{2} ({\bf r}) \exp \left[ i \phi \right],
\end{align}
and (\ref{eqn1-10}) becomes
\begin{align}
\label{eqn2-11}
& \overline{P}_{js,j's'}    =  2 \omega_{0}  \\ \nonumber
& \int_{{\cal V}_{1}} d^{3}{\bf r} \int_{{\cal V}_{1}} d^{3}{\bf r}' \, {\bf F}_{01,js}^{1 \ast} ({\bf r}) \cdot  \overline{\overline{\bm D}}^{\, 11} ({\bf r},{\bf r}',\omega_{0}) \cdot {\bf F}_{01,js}^{1} ({\bf r}') \\ \nonumber
& +  \int_{{\cal V}_{2}} d^{3}{\bf r} \int_{{\cal V}_{1}} d^{3}{\bf r}' \\ \nonumber
& \hspace{12mm} {\bf F}_{02,j's'}^{2 \ast} ({\bf r})  \cdot  \overline{\overline{\bm D}}^{\, 21} ({\bf r},{\bf r}',\omega_{0}) \cdot {\bf F}_{01,js}^{1} ({\bf r}')  \exp \left[ - i \phi \right] \\ \nonumber
& +  \int_{{\cal V}_{1}} d^{3}{\bf r} \int_{{\cal V}_{2}} d^{3}{\bf r}' \\ \nonumber
&  \hspace{12mm} {\bf F}_{01,js}^{1 \ast} ({\bf r}) \cdot  \overline{\overline{\bm D}}^{\, 12} ({\bf r},{\bf r}',\omega_{0}) \cdot {\bf F}_{02,j's'}^{2} ({\bf r}') \exp \left[ i \phi \right]  \\ \nonumber
& + \int_{{\cal V}_{2}} d^{3}{\bf r} \int_{{\cal V}_{2}} d^{3}{\bf r}' \, {\bf F}_{02,j's'}^{2 \ast} ({\bf r}) \cdot  \overline{\overline{\bm D}}^{\, 22} ({\bf r},{\bf r}',\omega_{0}) \cdot {\bf F}_{02,j's'}^{2} ({\bf r}').
\end{align}
The first and last terms are the powers dissipated by the two sources individually, into their respect loss mechanisms. The second and third terms lead to a fringe, which only exists when there is a cross coupling in the system. Notice the mixed domains on the integrals. The dissipated power can be written
\begin{align}
\label{eqn2-12}
\overline{P}^{12}_{js,j's'}  =  2 \omega_{0} \left[ D^{11}_{js,js}  \right. & +  D^{22}_{j's',j's'} \\ \nonumber
& \left. + 2 | D^{12}_{js,j's'} | \cos \left(  \phi + \theta_{js,j's'} \right) \right].
\end{align}
We have used the fact that the overall tensor $\overline{\overline{\bm D}} ({\bf r},{\bf r}',\omega_{0})$ is Hermitian, from which it follows that $\overline{\overline{\bm D}}^{\, 12} ({\bf r},{\bf r}',\omega_{0}) =  \overline{\overline{\bm D}}^{\, 21 \dagger} ({\bf r}',{\bf r},\omega_{0})$, which is Onsager's reciprocity \cite{ref16}. It follows that $D^{12}_{js,j's'} = D^{21 \ast}_{j's',js}$. The complex visibility of the observed fringe gives the real and imaginary parts of $D^{12}_{js,j's'}$, which are the matrix elements of the cross response tensor in the basis of the source fields:
\begin{align}
\label{eqn2-13}
D^{12}_{js,j's'}  = & \int_{{\cal V}_{1}} d^{3}{\bf r} \int_{{\cal V}_{2}} d^{3}{\bf r}' \\ \nonumber
 & {\bf F}_{01,js}^{1 \ast} ({\bf r}) \cdot  \overline{\overline{\bm D}}^{\, 12} ({\bf r},{\bf r}',\omega_{0}) \cdot {\bf F}_{02,j's'}^{2} ({\bf r}').
\end{align}
The matrix elements in this case are evaluated with respect to two different vector spaces, not least because ${\cal V}_{1}$ and ${\cal V}_{2}$ can be different.

The cross response tensor is then reconstructed through
\begin{equation}
\label{eqn2_14}
\overline{\overline{\bm D}}^{12} ({\bf r},{\bf r}',\omega) \approx \sum_{k,k'} D_{k,k'}^{12} \widetilde{\bf F}_{01,k}^{1} ({\bf r}) \widetilde{\bf F}_{02,k'}^{2 \ast}({\bf r}')
\mbox{.}
\end{equation}
where $\widetilde{\mathbb{F}}^{n}$ is the dual set of $\mathbb{F}^{n}$, over the appropriate domain.

In summary, interferometry can be used to find the matrix elements of the anti-Hermitian part of the generalized susceptibility tensor in the vector space of the field patterns of the applied forces. Dual functions can then be used to reconstruct the response tensor in the space domain. One may only be interested in the spatial correlations corresponding to one kind of force, in which case it is sufficient to carry out an experiment with two sources of the same kind; or one may be interested  in finding the spatial correlations corresponding to two kinds of force, in which case it is possible to use two different kinds of source. Two different sources create fringes that isolate and extract information relating to cross-correlated response.

\section{Response tensor decomposition}

What information is contained in the susceptibility tensor, and how many degrees of freedom need to be found?
The susceptibility tensor and force correlation tensor are, by definition, Hermitian when considered over all variables: position, polarisation, and type.  The response tensor only appears as the kernel of an integral equation (\ref{eqn1-10}), and so it is appropriate to look for a discrete decomposition. It can be shown that a tensor field, $\overline{\overline{\bf D}}^{\, mn} ({\bf r},{\bf r}')$, is Hilbert Schmidt \cite{ref17} if
\begin{align}
\label{eqn3-1}
\sum_{mn} \int_{{\cal V}_{m}} {\rm d}^{3} {\bf r}   \int_{{\cal V}_{n}}  {\rm d}^{3}{\bf r}'  \, \overline{\overline{\bf D}}^{\, mn} ({\bf r},{\bf r}') \cdot \cdot \,
\overline{\overline{\bf D}}^{\, nm \dagger} ({\bf r}',{\bf r}) & <  \infty
\mbox{.}
\end{align}

Every physical system must satisfy this condition. According to (\ref{eqn1-11}), $\overline{\overline{\bf D}}^{\, mn} ({\bf r},{\bf r}')$ comprises a forward and time-reversed process, both of which map a set of generalized forces onto a set of responsive perturbations. Because there is only a finite number of physical degrees of freedom available for effecting this mapping, (\ref{eqn3-1}) follows. Equivalently, the response tensor has a finite coherence volume, wherever it is measured, and the system occupies a finite region, and therefore there is a finite number of degrees of freedom available. A truly local response having the form
$\delta( {\bf r} - {\bf r}')$ is not physically possible because there would be an infinite number of degrees of freedom in every finite volume. A similar condition holds for the force correlation tensor:
\begin{align}
\label{eqn3-2}
\sum_{mn} \int_{{\cal V}_{m}} {\rm d}^{3}{\bf r}  \int_{{\cal V}_{n}}  {\rm d}^{3} {\bf r}' \, \overline{\overline{\bm F}}^{mn} ({\bf r},{\bf r}') \cdot \cdot \,
\overline{\overline{\bm F}}^{nm \dagger} ({\bf r}',{\bf r})  & <  \infty
\mbox{.}
\end{align}
Because any physical system must satisfy (\ref{eqn3-1}) and any realisable force must satisfy (\ref{eqn3-2}), the following Hilbert-Schmidt decompositions exist:
\begin{align}
\label{eqn3-3}
\overline{\overline{\bf D}}^{\, mn} ({\bf r},{\bf r}') & =   \sum_{i} \alpha_{i} {\bf d}^{m}_{i} ({\bf r}) {\bf d}_{i}^{n \ast} ({\bf r}') \\
\label{eqn3-4}
\overline{\overline{\bm F}}^{\, mn} ({\bf r},{\bf r}') & =  \sum_{j} \beta_{j} {\bf f}^{m}_{j} ({\bf r}) {\bf f}_{j}^{n \ast} ({\bf r}')
\mbox{.}
\end{align}
The basis set $\left\{{\bf d}^{m}_{i} ({\bf r}), \,  \forall i \in 1,\cdots, \infty  \right\}$ spans fields of type $m=1,2$ over the domains ${\cal V}_{1}$ and ${\cal V}_{2}$ respectively. The same is true of $\left\{{\bf f}^{m}_{i} ({\bf r}), \,  \forall i \in 1,\cdots, \infty  \right\}$. However, orthogonality is only guaranteed over the whole of the vector space, including the sum over $m$:
\begin{align}
\label{eqn3-5}
\sum_{m} \int_{{\cal V}_{m}} {\rm d}^{3}{\bf r} \, \, {\bf d}^{m}_{i} ({\bf r}) \cdot {\bf d}^{m \ast}_{j} ({\bf r}) = \delta_{ij}
\mbox{,}
\end{align}
which is undesirable in some circumstances, as will be discussed. The integrals in (\ref{eqn3-1}) can be evaluated by substituting (\ref{eqn3-3}) and using the orthogonality condition (\ref{eqn3-5}). This process gives $\sum_{i} \alpha_{i}^{2}$, and therefore (\ref{eqn3-1}) essentially states that the number of channels for absorbing power is limited. Likewise, (\ref{eqn3-2}) states that the number of channels available in the source that can do work is limited. For all systems, the eigenvalue spectrum, $\alpha_{i}$, tends rapidly to zero as some threshold value of $i$ is exceeded, and only a finite number of degrees of freedom need to be found when carrying out interferometry.

(\ref{eqn3-3}) and (\ref{eqn3-4}) can be substituted into (\ref{eqn1-12}) to give
\begin{equation}
\label{eqn3_6}
\langle \overline{P} \rangle  = 2 \omega \sum_{ij} \alpha_{i}  \beta_{j} \underbrace{\sum_{m} S_{ij}^{m} \sum_{n} S_{ij}^{n \ast}}_{\text{$t_{ij}$}},
\end{equation}
where
\begin{equation}
\label{eqn3_7}
S_{ij}^{m} =  \int_{{\cal V}_{m}}{\bf d}_{i}^{m} ({\bf r}) \cdot {\bf f}_{j}^{m \ast} ({\bf r})  \, {\rm d}^{3}{\bf r}
\mbox{.}
\end{equation}
(\ref{eqn3_6}) describes power absorption in terms of a scattering process, $t_{ij}$, that projects the natural modes of the forces, having weightings $\beta_{j}$, onto the natural modes of the system, having responsivities $\alpha_{i}$. When the system is driven by an incoherent superposition of its natural modes
$t_{ij} = \delta_{ij}$, and the system is maximally responsive with respect to spatial variations in the force.

(\ref{eqn3-3}) and (\ref{eqn3-4}), where a single set of basis functions spans both domains, are the most suitable decompositions in many cases. For example, if $m=1$ corresponds to an electric field and $m=2$ to a magnetic field, then these would be correlated if an electromagnetic wave is incident on the system. In this case, the diagonal block terms $m \neq n$ should be retained in $\overline{\overline{\bm F}}^{\, mn} ({\bf r},{\bf r}')$. Alternatively, the impressed field may, for example, comprise a physical force and magnetic vector potential, in which case the two generalized forces can be regarded as independent, and the block off-diagonals are not needed. Later we shall discuss the situation where one force is a scalar and the other a vector, as in the case of the electric scalar potential and magnetic vector potential.

Rather than using (\ref{eqn3-3}) and (\ref{eqn3-4}), there is a different approach, which seems better suited to decomposing data when only part of the susceptibility tensor is measured. Because (\ref{eqn3-1}) and (\ref{eqn3-2}) hold, the individual terms under the sum must also be Hilbert Schmidt, and because the block diagonal terms $m=n$ are each Hermitian, they can be diagonalised separately:
\begin{align}
\label{eqn3-8}
\overline{\overline{\bf D}}^{\, mm} ({\bf r},{\bf r}') & =   \sum_{i} \alpha_{i}^{m} {\bf d}^{m}_{i} ({\bf r}) {\bf d}_{i}^{m \ast} ({\bf r}') \\
\label{eqn3-9}
\overline{\overline{\bm F}}^{\, mm} ({\bf r},{\bf r}') & =  \sum_{j} \beta_{j}^{m} {\bf f}^{m}_{j} ({\bf r}) {\bf f}_{j}^{m \ast} ({\bf r}')
\mbox{.}
\end{align}
In this case, $\left\{ {\bf d}^{m}_{i} ({\bf r}), \forall i \in 1,\cdots, \infty \right\}$ forms a complete orthonormal basis over ${\cal V}_{m}$. Different orthogonal basis sets are therefore generated for the two domains. The same is true of the force basis $\left\{ {\bf f}^{m}_{i} ({\bf r}), \forall i \in 1,\cdots, \infty \right\} $.

Consider what happens when only one kind of force is present, say $m=n$. Substituting (\ref{eqn3-8}) and (\ref{eqn3-9}) into (\ref{eqn1-12}) gives
\begin{equation}
\label{eqn3_10}
\langle \overline{P} \rangle = 2 \omega \sum_{ij} \alpha_{i}^{m}  \beta_{j}^{m}   | S_{ij}^{m} |^{2}.
\end{equation}
(\ref{eqn3-9}) describes the partially coherent generalized force in terms of an incoherent superposition of fully coherent fields, with weighting factors $\beta_{j}^{m}$. These are the natural modes of the illumination, as introduced in the context of optics by Wolf \cite{ref14}. (\ref{eqn3-8}) describes the absorptive response in terms of a set of orthogonal modes, each having responsivity $\alpha_{i}^{m}$. According to (\ref{eqn3_10}), the natural modes of the force scatter, with efficiencies $|S_{ij}^{m}|^{2}$, into the modes to which the system is responsive. This representation constitutes the coupled-mode model \cite{ref4a,ref4b} of power absorption. Again, maximum coupling is achieved when the modes of the field match those of the system, over the appropriate domain, which defines the state of coherence to which the system is maximally receptive as the spatial form of the impressed field is varied.

If the system responds in an entirely local way, $\overline{\overline{\bm D}}^{mm} ({\bf r},{\bf r}') =
\overline{\overline{\bm D}}_{0}^{mm}({\bf r}) \delta({\bf r} - {\bf r}')$, a Hilbert Schmidt decomposition does not exist, but (\ref{eqn1-12}), still results in finite power, because the number of channels available for absorbing power is limited by the smoothness of the impressed force. Because, in this case, the natural modes of the system span any force distribution over ${\cal V}_{m}$, it behaves as a near-perfect absorber; the generalized equivalent of a `light bucket'.

Now consider the case where two different kinds of force are present simultaneously. In order to calculate the absorbed power, it is necessary to calculate the Hilbert Schmidt decomposition of the cross terms $m \neq n$:
\begin{align}
\label{eqn3-11}
\overline{\overline{\bf D}}^{\, mn} ({\bf r},{\bf r}') & =   \sum_{i} \alpha_{i}^{mn} {\bf d}^{m \prime}_{i} ({\bf r}) {\bf d}_{i}^{n \prime \ast} ({\bf r}') \\
\label{eqn3-12}
\overline{\overline{\bm F}}^{\, mn} ({\bf r},{\bf r}') & =  \sum_{j} \beta_{j}^{mn} {\bf f}^{m \prime}_{j} ({\bf r}) {\bf f}_{j}^{n \prime \ast} ({\bf r}')
\mbox{.}
\end{align}
Primes have been used to indicate that the natural basis functions that describe the cross response may be different to those that describe the self response. (\ref{eqn3-11}) and (\ref{eqn3-12}) have the forms needed to ensure that the overall response tensor is Hermitian. (\ref{eqn3-12}) describes the cross correlations in terms of an incoherent superposition of fully coherent field pairs. In other words for every basis function in domain ${\cal V}_{1}$, ${\bf f}^{1 \prime}_{j} ({\bf r})$, there is a unique, associated basis function in ${\cal V}_{2}$, ${\bf f}^{2 \prime}_{j} ({\bf r})$. These basis functions are in one-to-one correspondence, revealing generalized force distributions in the two domains that are mutually fully coherent and uniquely related. For example, one might correspond to an electric field and the other to a magnetic field. The Hilbert-Schmidt decomposition of the off-diagonal block $n \neq m$, therefore describes the cross correlations between two different vector spaces as a weighted linear combination of field pairs. This approach generalizes Wolf's formalism to include cross correlations between different vector spaces.

The same decomposition can be carried out on the susceptibility tensor. The reason for decomposing the on-diagonal and off-diagonal blocks individually, is that it is only necessary to carry out partial interferometric measurements---say using two sources of the first kind, or two of the second kind, or one of each---in order to reveal collective behaviour. It should also be appreciated that the force can be described in terms of one scheme, say (\ref{eqn3-4}), and the system in terms of the other, say (\ref{eqn3-8}) and (\ref{eqn3-11}), and (\ref{eqn1-12}) still returns the correct result for the absorbed power.

\begin{figure}[h]
\noindent \begin{centering}
\includegraphics[trim = 3cm 17cm 7cm 4cm,width=80mm]{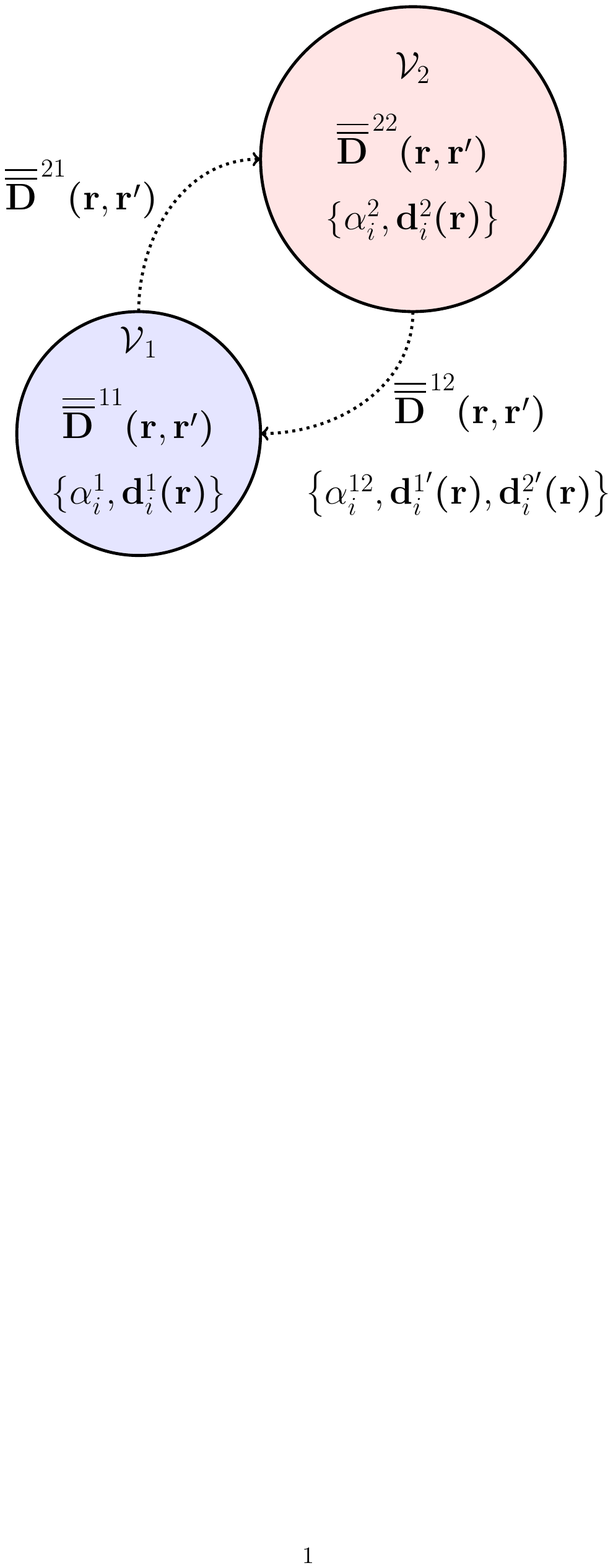}
\par\end{centering}
\caption{\small Two domains corresponding to two different generalized forces. They may be fully separated, partially overlapping, or the same. The dissipative response in domain ${\cal V}_{1}$ is spanned by the orthonormal collective excitations ${\bf d}^{1}_{i} ({\bf r})$ having responsivties $\alpha_{i}^{1}$. The dissipative response in domain ${\cal V}_{2}$ is spanned by the orthonormal collective excitations ${\bf d}^{2}_{i} ({\bf r})$ having responsivities $\alpha_{i}^{2}$. The cross-correlated response between the two domains is described by the function pairs ${\bf d}^{1 \prime}_{i} ({\bf r})$ and ${\bf d}_{i}^{2 \prime \ast} ({\bf r}')$ with relative weightings $\alpha_{i}^{12}$.}
\label{fig2}
\end{figure}

The process of decomposing the self- and cross-subspaces can be summarised as follows:
\begin{align}
\label{eqn3-13}
\left[
\begin{array}{c|c}
  \sum_{i} \alpha_{i}^{1} {\bf d}^{1}_{i} ({\bf r}) {\bf d}_{i}^{1 \ast} ({\bf r}')  & \sum_{i} \alpha_{i}^{12} {\bf d}^{1 \prime}_{i} ({\bf r}) {\bf d}_{i}^{2 \prime \ast} ({\bf r}')  \\ \hline
  \sum_{i} \alpha_{i}^{21} {\bf d}^{2 \prime }_{i} ({\bf r}') {\bf d}_{i}^{1 \prime \ast} ({\bf r})  & \sum_{i} \alpha_{i}^{2} {\bf d}^{2}_{i} ({\bf r}) {\bf d}_{i}^{2 \ast} ({\bf r}')
\end{array}
\right]
\mbox{,}
\end{align}
which is shown schematically in Figure~\ref{fig2}.

The top left block in (\ref{eqn3-13}) corresponds to the decomposition obtained when interferometric measurements are made using sources of type 1 only. The Hilbert-Schmidt decomposition, which is a diagonalisation in this case, gives the individual natural modes through which the structure can absorb power from a partially coherent force of type 1; the eigenvalues are the associated responsivities.  In addition, the bottom right block can be measured and decomposed in the same way, giving a full description of the system's ability to absorb power from a partially coherent source of type 2. If only the off-diagonal blocks in (\ref{eqn3-13}) are measured, the Hilbert-Schmidt decomposition describes cross-correlated response.

An interesting question is how do the functions ${\bf d}^{m}_{i} ({\bf r})$ and ${\bf d}^{m \prime }_{i} ({\bf r})$ relate to each other? Ordinarily it might be expected that the modes responsible for absorbing power from the sources individually are the same as the modes responsible for modulating the behaviour when two different kinds of force are applied simultaneously: in other words, ${\bf d}^{m}_{i} ({\bf r})$ and ${\bf d}^{m \prime }_{i} ({\bf r})$ are the same. The real elements on the leading diagonal of the whole tensor account for all energy dissipation mechanisms present. The off-diagonal blocks account for work done by one kind of force on the conjugate variable associated with the other source, and therefore account for the modulation in the dissipated power when sources of two kinds are present. They do not represent power dissipation mechanisms in their own right, and certainly, the absorbed power, given by (\ref{eqn2-8}), cannot become negative as the phase between the interferometric sources is varied.

$\overline{\overline{\bf D}}^{\, 11} ({\bf r},{\bf r}')$ and $\overline{\overline{\bf D}}^{\, 22} ({\bf r},{\bf r}')$ have null spaces corresponding to those force distributions that cannot dissipate power in the system. In other words the $\alpha_{i}^{m}$ tend rapidly to zero above some critical index $I^{m}_{c}$. Likewise the $\alpha_{i}^{12}$ tend rapidly to zero above some critical index $I_{c}^{12}$. The null spaces of the off-diagonal blocks span, at least, the null spaces of the diagonal blocks, and $I^{m}_{c} \ge I^{12}_{c}$, which can lessen the amount of experimental work needed if the whole tensor is measured. In fact, the sampling strategy can be chosen to ensure that any cross-correlations present will be found. The natural basis functions, ${\bf d}^{m}_{i} ({\bf r})$ and ${\bf d}^{m \prime }_{i} ({\bf r})$, do not have to be the same, but the  ${\bf d}^{m}_{i} ({\bf r})$ must span the ${\bf d}^{m \prime }_{i} ({\bf r})$. The cross-correlated response can be described in terms of the modes of the self correlations. Ultimately, the precise relationship between the decompositions depends on the nature of the physical system being studied. To keep the analysis general, we prefer to calculate the natural modes in the two domains on the basis of the diagonal blocks, giving ${\bf d}_{i}^{m} ({\bf r})$, and then to project the natural modes of the cross correlations, ${\bf d}^{m \prime}_{i} ({\bf r})$ onto those basis sets to look for spatial relationships between the self and cross correlations.

To this point, we have assumed that both generalized forces are vector fields, but consider what happens when one force is a scalar and the other a vector. The overall generalized force is then described by a four-vector. In the case of an electric scalar potential and a magnetic vector potential, the use is clear. For any general four vector, the block decomposition becomes
\begin{align}
\label{eqn3-14}
\left[
\begin{array}{c|c}
  \sum_{i} \alpha_{i}^{1} d_{i} ({\bf r}) d_{i} ({\bf r}')  & \sum_{i} \alpha_{i}^{12} d^{\prime}_{i} ({\bf r}) {\bf d}_{i}^{\prime \ast} ({\bf r}')  \\ \hline
  \sum_{i} \alpha_{i}^{21} {\bf d}^{\prime }_{i} ({\bf r}') d_{i}^{\prime \ast} ({\bf r})  & \sum_{i} \alpha_{i}^{2} {\bf d}_{i} ({\bf r}) {\bf d}_{i}^{\ast} ({\bf r}')
\end{array}
\right]
\mbox{.}
\end{align}
The top left block, which is spanned by the scalar functions $\left\{ d_{i} ({\bf r}), \forall i \in 1,\cdots, \infty \right\}$ over the domain ${\cal V}_{1}$, completely characterises the response to the scalar force alone. The bottom right block, which is spanned by the vector functions $\left\{ {\bf d}_{i} ({\bf r}), \forall i \in 1,\cdots, \infty \right\}$ over the domain ${\cal V}_{2}$, completely characterises
the response to the vector force alone. The off-diagonal blocks describe spatial cross correlations between the scalar and vector fields. In other words, there are certain scalar fields that map in one-to-one correspondence with certain vector fields, and these characterise the spatial forms of the interactions in the system.

\section{Scattering}

It is common practice to describe microscopic solid-state behaviour using quantum correlation functions, and then to wrap the solid-state behaviour in a classical scattering model to describe macroscopic behaviour.  Often, the quantum correlation function is determined for an infinitely large system, and the boundary effects of a real sample are introduced through scattering. For example, the dielectric properties of a material may be calculated by using Kubo's formula, and then the susceptibility used in an electromagnetic model based on Maxwell's equations \cite{ref18}. For physically small systems, this distinction is not possible. Ultimately, the boundary between the two regimes depends on which interactions are included in the Hamiltonian. Another example is when classical dipolar interactions are used as the mediating force in spin waves, but the individual precessing elements are quantised.

In the context of interferometry, scattering is important because it determines the degree to which one can gain access to the intrinsic properties of a material. For example, when measuring the intrinsic properties of a magnetic material it is desirable to correct for the demagnetisation field; or when trying to determine the bulk electromagnetic properties of a material, it is necessary to correct for skin depth. Similar issues arise in acoustics, where the boundary conditions at the edges of the sample must be included.

In the case of sensors, there is an important relationship between the coherence length of the intrinsic solid-state absorption mechanism and the physical size of the absorber in determining the number and form of the degrees of freedom available for absorbing power. As the physical size approaches the intrinsic coherence length, the number of degrees of freedom and their individual efficiencies decrease rapidly, which has implications for many applications such as far-infrared sensors and near-field radiative heat transfer between nano-scale structures.

In order not to hide the central message, we will consider scattering in the case where only one type of force is present, but the extension to two forces is straightforward. Suppose that the generalized force at ${\bf r}$ has two parts:
\begin{align}
\label{eqn4-1}
{\bf F} ({\bf r}) = {\bf F}^{\rm e} ({\bf r}) + {\bf F}^{\rm s} ({\bf r})
\mbox{.}
\end{align}
${\bf F}^{\rm e} ({\bf r})$ is the applied external force, and ${\bf F}^{\rm s} ({\bf r})$ is the additional generalized force that results from scattering. Here we shall assume that scattering occurs as a result of the perturbation of the system itself, but situations where scattering occurs as a consequence of some external body can be covered by the same formalism.

The scattered field, which is the field re-radiated by excited elements in the system,  is a linear function of the external field, and therefore the total field is a linear function of the external field:
\begin{equation}
\label{eqn4-2}
{\bf F} ({\bf r})  =   \int_{\cal V} {\rm d}^{3} {\bf r}' \, \overline{\overline{\bm G}} ({\bf r},{\bf r}') \cdot {\bf F}^{\rm e} ({\bf r}')
\mbox{,}
\end{equation}
where $\overline{\overline{\bm G}} ({\bf r},{\bf r}')$ is the appropriate dyadic scattering kernel. The integral is taken over the volume of the sample, and so  the scattering kernel does not take into account the propagation of the impressed field from the source position to the sample. $\overline{\overline{\bm G}} ({\bf r},{\bf r}')$ is intrinsic to the sample, accounting for sample-dependent effects. (\ref{eqn4-2}) is quite general, and the appropriate operator can be found by either analytical or numerical means. In electromagnetism it gives rise to the so called Electric Field Integral Equation (EFIE) and the Magnetic Field Integral Equation(MFIE) \cite{ref18}.

Using the total field from (\ref{eqn4-2}) in (\ref{eqn1-10}), gives
\begin{align}
\label{eqn4-3}
\langle \overline{P} \rangle = & 2 \omega \int_{\cal V}  {\rm d}^{3} {\bf r}' \int_{\cal V}  {\rm d}^{3}{\bf r} \\ \nonumber
& \overline{\overline{\bf K}} ({\bf r},{\bf r}',\omega) \cdot \cdot \,
\overline{\overline{\bf F}}^{{\rm e} \dagger} ({\bf r}',{\bf r},\omega)
\mbox{,}
\end{align}
where
\begin{align}
\label{eqn4-4}
\overline{\overline{\bf K}} ({\bf r},{\bf r}',\omega) = & \int_{\cal V}  \, {\rm d}^{3} {\bf s}' \int_{\cal V}  \, {\rm d}^{3}{\bf s} \\ \nonumber
& \overline{\overline{\bf G}}^{\dagger} ({\bf r},{\bf s},\omega) \cdot \overline{\overline{\bf D}} ({\bf s},{\bf s}',\omega) \cdot \overline{\overline{\bf G}} ({\bf s}',{\bf r}',\omega)
\mbox{,}
\end{align}
is the response dyadic of the complete sample. We call $\overline{\overline{\bf K}} ({\bf r},{\bf r}',\omega)$ the response dyadic of the sample, because it describes the state of coherence of the field to which the whole sample is sensitive.  The natural modes of $\overline{\overline{\bf K}} ({\bf r},{\bf r}',\omega)$ give the amplitude, phase and polarisation patterns of the individual channels through which the whole sample can absorb energy from a generalized force, taking into account internal scattering. Often, say in the case of a detector, $\overline{\overline{\bf K}} ({\bf r},{\bf r}',\omega)$ is all that is needed in order to characterise behaviour.

Equation~(\ref{eqn4-4}) shows that $\overline{\overline{\bf G}} ({\bf s}',{\bf r}',\omega)$ acts as a `filter' that wraps around the intrinsic response, and which limits the amount of spatial information available. This interpretation follows because $\overline{\overline{\bf G}} ({\bf s}',{\bf r}',\omega)$ is a operator having a finite throughput. It limits the degrees of freedom available for absorbing power, and equally it limits the information that can be determined about the intrinsic absorption mechanism described by $\overline{\overline{\bf D}} ({\bf s},{\bf s}',\omega)$.

$\overline{\overline{\bf G}} ({\bf s},{\bf r},\omega)$ contains the full susceptibility dyadic, including the non-dissipative part, and rarely can it be `deconvolved' from a measurement. For example, if one or both of the the points ${\bf s}$ and ${\bf s}'$, in  $\overline{\overline{\bf D}} ({\bf s},{\bf s}',\omega)$, is deep inside a sample, deeper than the skin depth, it is not possible to determine information about the deep spatial structure by carrying out external measurements. The degree to which screening is an obstacle depends on what the experimenter wants to achieve. Sometimes, say in the case of detector characterization, it is sufficient to know the response dyadic of the overall sample, and only the parts that are accessible to external influences can contribute to the absorption process, and nothing else is of importance. Sometimes the objective is to measure the intrinsic non-local susceptibility of the material, and then the sample geometry and scanning strategy must be chosen in accordance with the particular need.

\section{K-domain formulation}

When calculating quantum correlation functions, it is common practice to cast the Hamiltonian into the wave-vector, ${\bf k}$, domain; for example, potential functions are often easier to describe in the ${\bf k}$-domain than in the space domain. One advantage is that for translationally invariant systems, which require the system to have infinite extent, the response function can be written solely in terms of a single ${\bf k}$ variable. As discussed, it is possible, often desirable, to calculate the intrinsic properties of a system having infinite extent, and then to take into account the finite size of the actual sample through a classical scattering analysis \cite{ref18}. The important point is that response functions are often expressed in the ${\bf k}$-domain and so it is desirable to describe EAI in the ${\bf k}$-domain.

\begin{figure}[h]
\noindent \begin{centering}
\includegraphics[trim = 2.0cm 17cm 9cm 3.5cm,width=80mm]{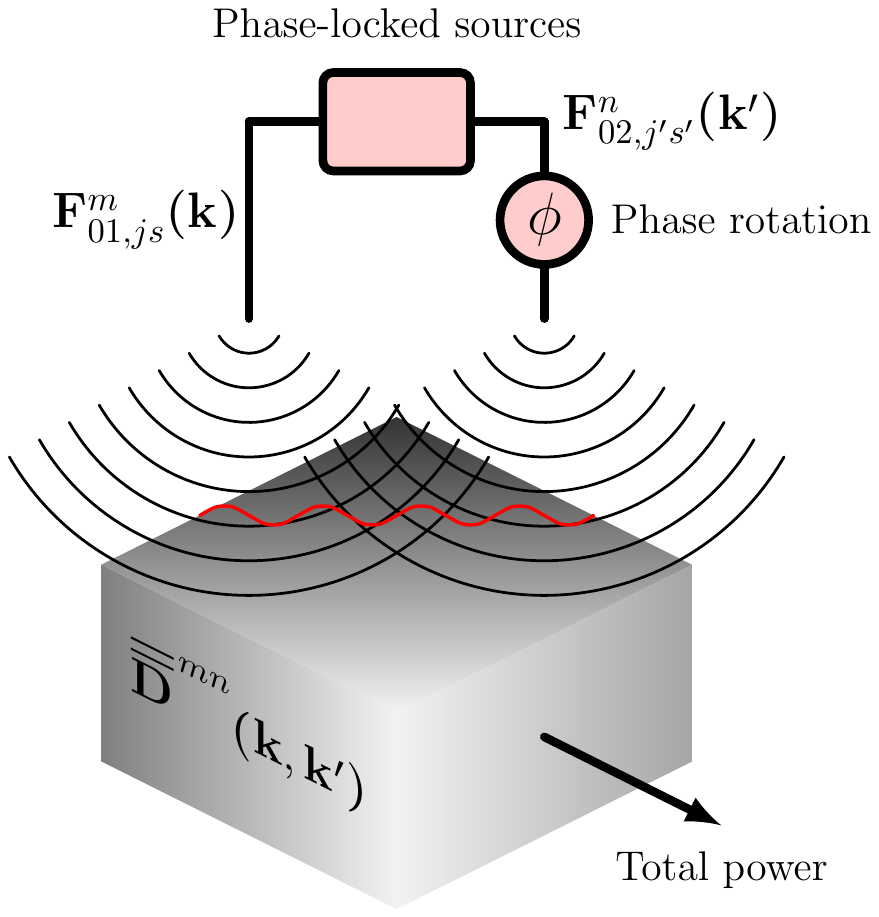}
\par\end{centering}
\caption{\small Energy Absorption Interferometer. Two phase-locked sources produce generalized forces ${\bf F}_{01,js}^{m} ({\bf k})$ and  ${\bf F}_{02,j's'}^{n} ({\bf k}')$. The time-averaged total power displays a fringe as the differential phase $\phi$ is varied. The complex visibilities for different source locations enable the complex-valued system response tensor $\overline{\overline{\bf D}}^{\, mn} ({\bf k},{\bf k}')$ to be determined in the $\bf k$ domain.}
\label{fig4}
\end{figure}

We shall use the following  ${\bf k}$-domain representation of the response tensor:
\begin{align}
\label{eqn5-1}
\overline{\overline{\bf D}}^{\, mn} ({\bf r}, {\bf r}',\omega) & = \frac{1}{(2 \pi)^{3}} \int_{-\infty}^{+\infty} d^{3}{\bf k} \frac{1}{(2 \pi)^{3}}  \int_{-\infty}^{+\infty} d^{3} {\bf k}' \\ \nonumber
& \overline{\overline{\bm D}}^{\, mn} ({\bf k}, {\bf k}',\omega)  \exp \left[- i {\bf k}\cdot{\bf r} \right] \exp \left[ + i {\bf k}' \cdot {\bf r}' \right],
\end{align}
and a similar expression for the force correlation tensor
\begin{align}
\label{eqn5-2}
\overline{\overline{\bf F}}^{\, mn} ({\bf r}, {\bf r}',\omega) & = \frac{1}{(2 \pi)^{3}} \int_{-\infty}^{+\infty} d^{3}{\bf k} \frac{1}{(2 \pi)^{3}}  \int_{-\infty}^{+\infty} d^{3} {\bf k}' \\ \nonumber
& \overline{\overline{\bm F}}^{\, mn} ({\bf k}, {\bf k}',\omega)  \exp \left[- i {\bf k}\cdot{\bf r} \right] \exp \left[ + i {\bf k}' \cdot {\bf r}' \right].
\end{align}

Substituting (\ref{eqn5-1}) and (\ref{eqn5-2}) in (\ref{eqn1-12}) gives
\begin{align}
\label{eqn5-3}
\langle \overline{P} \rangle   = & 2 \omega \sum_{mn} \frac{1}{(2 \pi)^{3}}  \int_{\infty}^{+\infty} d^{3}{\bf k} \frac{1}{(2 \pi)^{3}}  \int_{\infty}^{+\infty} d^{3}{\bf k}' \\ \nonumber
& \overline{\overline{\bm D}}^{\, mn} ({\bf k},{\bf k}',\omega) \cdot \cdot \, \overline{\overline{\bf F}}^{nm \dagger} ({\bf k}',{\bf k},\omega).
\end{align}
The total power absorbed again takes the form of the contraction of two tensor fields to a scalar, but now the contraction is carried out in the ${\bf k}$-domain.

In the case of an interferometric measurement, Figure~\ref{fig4}, where the sources are fully coherent, (\ref{eqn5-3}) can be written
\begin{align}
\label{eqn5-4}
\overline{P}   = & 2 \omega_{o} \sum_{mn} \frac{1}{(2 \pi)^{3}}  \int_{\infty}^{+\infty} d^{3}{\bf k} \frac{1}{(2 \pi)^{3}}  \int_{\infty}^{+\infty} d^{3}{\bf k}' \\ \nonumber
& {\bf F}^{m \ast} ({\bf k},\omega_{0}) \cdot
\overline{\overline{\bm D}}^{\, mn} ({\bf k},{\bf k}',\omega_{0})
\cdot {\bf F}^{n} ({\bf k}',\omega_{0}).
\end{align}
The matrix elements of the response tensor are now calculated in the $\bf k$ domain. In those cases where the sources produce plane waves, an interferometric measurement records specific elements of the ${\bf k}$-domain response tensor directly. If the Hamiltonian is shift invariant, the response tensor has the form $\overline{\overline{\bm D}}^{mn} ({\bf k},{\bf k}',\omega_{0}) = (2 \pi )^{3} \overline{\overline{\bm D}}_{0}^{mn}({\bf k},\omega_{0}) \delta({\bf k} - {\bf k}')$, which indicates an infinitely small correlation angle, and  (\ref{eqn5-4}) becomes
\begin{align}
\label{eqn5-5}
\overline{P}    = & 2 \omega_{0} \sum_{mn} \frac{1}{(2 \pi)^{3}}  \int_{\infty}^{+\infty} d^{3}{\bf k}\\ \nonumber
& {\bf F}^{m \ast} ({\bf k},\omega_{0}) \cdot \overline{\overline{\bm D}}^{\, mn}_{0} ({\bf k},\omega_{0})
\cdot {\bf F}^{n} ({\bf k},\omega_{0}).
\end{align}

Only a single source is needed to scan the angular response. All structures have finite size, however, and regardless of whether this is included in the Hamiltonian or as a classical scattering process, the effect is to create an angular response having a finite coherence angle. In the case of plane waves, far-field measurements, the smallest feature that can be resolved is determined by the wavelength of the impressed field, which together with restrictions on the polarisation, effectively induce angular correlations in the response tensor of the field \cite{ref19a,ref19b} and of the sample. We shall not elaborate on these issues here. In the context of sensors, the interferometric method measures the far-field angular response tensor, which can then be decomposed to give the amplitude, phase, polarisation patterns, and responsivities of the individual fully-coherent `antenna patterns' through which the sample can absorb power.

The ${\bf k}$-domain formulation is similar to aperture synthesis interferometry used in astronomy \cite{ref20}, but the process is carried out in absorption rather than in emission. In astronomy, the Fourier components of the sky brightness distribution are measured, using the fact that the projected relative separation between the telescopes varies as the Earth rotates. Great care is taken to ensure that the telescopes are positioned in such a way that the Fourier components are sampled fully for a given class of source. In EAI, each pair of sources impresses a certain Fourier field having some orientation. As the differential phase is varied this impressed Fourier modulation shifts along its length. One is therefore measuring the real and imaginary parts of a Fourier component of the system's ability to absorb energy. Numerous elegant experimental and data processing techniques have been developed for aperture synthesis astronomy, and we believe that many of these these can be adapted to energy absorption interferometry.

\section{Data analysis}

It is valuable to describe a numerical implementation of the proposed scheme, both from the perspective of simulating data and also from the perspective of analysing results. The system of interest may intrinsically comprise a collection of discrete elements, or it may be divided into sample volumes, in much the same way as the Discrete Dipole Approximation \cite{ref21}. Let $\mathsf{f}^{{\rm e} m} \in \mathbb{C}^{3J}$ be a column vector containing the complex amplitudes of the Cartesian components of applied force $m$ at the positions of the sample points. If two kinds of force are present simultaneously, the absorbed power is given by
\begin{align}
\label{eqn6-1}
 P   & =   2 \omega_{0}  \sum_{mn}  \mathsf{f}^{{\rm e} m \dagger} \mathsf{D}^{mn} \, \mathsf{f}^{{\rm e} n}   \\ \nonumber
& =   2 \omega_{0}  \sum_{mn} {\rm Tr} \left[ \mathsf{f}^{{\rm e} m \dagger} \mathsf{D}^{mn} \, \mathsf{f}^{{\rm e} n}  \right] \\ \nonumber
\langle P \rangle & = 2 \omega_{0}  \sum_{mn} {\rm Tr} \left[ \mathsf{D}^{mn} \mathsf{N}^{nm \dagger} \right]
\mbox{.}
\end{align}
$\mathsf{N}^{nm \dagger} = \langle \mathsf{f}^{{\rm e} n} \mathsf{f}^{{\rm e} m \dagger} \rangle \in \mathbb{C}^{3J \times 3J}$ is the $n,m$'th block of the adjoint of the spatial correlation matrix of the applied forces, and $\mathsf{D}^{mn} \in \mathbb{C}^{3J \times 3J}$ is the $m,n$'th block of the response matrix of the sample.

For brevity, it is convenient to assume that only one kind of force is present $m=n=1$, but the extension to two forces is straightforward. Omitting the superscripts gives
\begin{align}
\label{eqn6-2}
P  & =   2 \omega_{0}  {\rm Tr} \left[ \mathsf{D} \mathsf{N}^{\dagger} \right]
\mbox{,}
\end{align}
where $D$ and $N$ are Hermitian.

The Trace of the product of a matrix with the adjoint of another is an inner product in the abstract vector space of complex Hermitian matrices, and so (\ref{eqn6-2}) projects the state of coherence of the force onto the state of coherence to which is system is maximally sensitive. If scattering is included then $\mathsf{D}$ should be replaced by the overall response matrix $\mathsf{K} = \mathsf{G}^{\dagger} \mathsf{D} \mathsf{G}$,
where $ \mathsf{G}$ is the scattering matrix, which is the discretised version of (\ref{eqn4-4}). $\mathsf{D}$ and $\mathsf{K}$ are Hermitian and can be diagonalised to give the natural modes through which the structure can absorb energy. The eigenvectors correspond to coherent dynamical modes of the system, and the eigenvalues to their responsivities, and therefore interferometry can be used to uncover the individual coherent excitations.

Consider applying interferometry to Eq.~(\ref{eqn6-2}). Assemble a matrix $\mathsf{F}^{\rm src} \in \mathbb{C}^{2J \times N}$ where each column contains the sampled force $\mathsf{f}^{\rm e}$ associated with some particular source position and orientation. The number of source positions and orientations is $N$, and
the minimum number needed to allow the source distributions to be `deconvolved' from the data is $2J$. Fewer may be sufficient if interactions reduce the number of modes available for absorbing energy, or more may be used to over-sample the experiment.

If the columns of $\widetilde{\sf F}^{\rm src}$ contain the dual vectors to the columns of $\mathsf{F}^{\rm src}$ then by definition ${\sf F}^{{\rm src} \dagger} \widetilde{\sf F}^{\rm src} = {\sf I} =  \widetilde{\sf F}^{{\rm src} \dagger} {\sf F}^{\rm src}$, where ${\sf I}$ is the identity matrix. The dual vectors span the same vector space as the the original source vectors, and so it is possible to represent the response matrix in terms of the dual-vector basis:
\begin{equation}
\label{eqn6-3}
\mathsf{D}  =  \sum_{nn'} M_{nn'} \widetilde{\mathsf{f}}_{n} \widetilde{\mathsf{f}}^{\dagger}_{n'}  =  \widetilde{\mathsf{F}}^{\rm src} \mathsf{M} \widetilde{\mathsf{F}}^{{\rm src} \dagger}
\mbox{,}
\end{equation}
where the matrix elements of $\mathsf{D}$, in the dual basis, are contained in $\mathsf{M}$, and $n,n' \in  1 \cdots N$.

Consider a single interferometric measurement, where two phased-locked sources are present at $m$ and $m'$ with a variable phase difference, $\phi$, between them. Using (\ref{eqn6-1}) and (\ref{eqn6-3}), the absorbed power is given by
\begin{align}
\label{eqn6-4}
P_{mm'}  & =   \sum_{nn'} M_{nn'}  \Big[ \mathsf{f}^{\dagger}_{m} +  \mathsf{f}^{\dagger}_{m'} e^{-i \phi} \Big] \widetilde{\mathsf{f}}_{n} \widetilde{\mathsf{f}}^{\dagger}_{n'} \Big[ \mathsf{f}_{m} + \mathsf{f}_{m'} e^{i \phi} \Big] \\ \nonumber
 & =   M_{mm} + M_{m'm}  e^{-i \phi} + M_{mm'}  e^{i \phi} + M_{m'm'} \\ \nonumber
 & =   M_{mm} + M_{m'm'} + 2 | M_{m m'} | \cos \left( \phi + \phi_{m m'} \right)
\mbox{,}
\end{align}
where in the second line we have used the biorthogonality of the source vectors with their duals. $| M_{m m'} |$ and $\phi_{m m'}$ are the amplitude and phase of $ M_{m m'}$ respectively. If only source $m$ is turned on, the measured power is $M_{mm}$. If only source $m'$ is turned on, the measured power is $M_{m'm'}$. If both sources are turned on, and the differential phase varied, the power displays a fringe, which gives the real and imaginary parts of $M_{mm'} = M_{m'm}^{\ast}$. Thus the measured fringe reveals elements of the response matrix in the dual basis,
\begin{equation}
\label{eqn6-5}
\mathsf{M}  = \mathsf{F}^{\rm src \dagger} \mathsf{D} \mathsf{F}^{\rm src}
\mbox{.}
\end{equation}
$\mathsf{M}$ can be populated experimentally, and then the response matrix $\mathsf{D}$ calculated using (\ref{eqn6-3}). The use of dual vectors in Eq.~(\ref{eqn6-3}) essentially `deconvolves' the forms of the source fields from the measurement, over those regions that are accessible by the sources.

Dual vectors are required to calculate the response tensor in the space domain, and by definition ${\sf F}^{{\rm src} \dagger} \widetilde{\sf F}^{\rm src} = {\sf I}=  \widetilde{\sf F}^{{\rm src} \dagger} {\sf F}^{\rm src}$. Generally $J \neq N$, and it is not possible to invert ${\sf F}^{\rm src}$ directly. It is possible to calculate the pseudo-inverse through Singular Value Decomposition (SVD). If ${\sf F}^{\rm src} = {\sf U} {\sf \Sigma} {\sf V}^{\dagger}$, where ${\sf \Sigma}$ is diagonal, then $\widetilde{\sf F}^{\rm src} = {\sf U} {\sf \Sigma}^{-1} {\sf V}^{\dagger}$. The psuedo-inverse correctly takes into account whether there are more source positions than sample points or more sample points than source positions.

In what follows, assume that the interferometer uses point sources, which could for example take the form of electric or magnetic dipoles. Create the diagonal matrix $\mathsf{L}^{\rm src} \in {\cal C}^{2N} \times {\cal C}^{2N}$, where the diagonal entries are the complex amplitudes of the dipole moments of the sources. Strictly, $\mathsf{L}^{\rm src}$ has zero diagonal entries if not all of the available sample positions and orientations are used. Then,
\begin{equation}
\label{eqn6-6}
\mathsf{F}^{\rm src} = \mathsf{G}^{\rm src} \mathsf{L}^{\rm src}
\mbox{,}
\end{equation}
where each column of $\mathsf{G}^{\rm src}$ is the discretised source Green's function. $\mathsf{G}^{\rm src}$ maps the complex amplitude of the dipole moment of every possible source onto the vector components of the force at each sample point in the system. In this case
\begin{align}
\label{eqn6-7}
\mathsf{M}  & =  \mathsf{L}^{{\rm src} \dagger} \mathsf{G}^{{\rm src} \dagger} \mathsf{D} \mathsf{G}^{\rm src} \mathsf{L}^{\rm src} \\ \nonumber
 & = \mathsf{L}^{{\rm src} \dagger} \mathsf{M}' \mathsf{L}^{\rm src}
\mbox{.}
\end{align}
The response matrix $\mathsf{M}' = \mathsf{G}^{{\rm src} \dagger} \mathsf{D} \mathsf{G}^{\rm src}$ characterises the resonse of the system in terms of point sources at the positions of the sources. It can be diagonalised to give the natural modes referenced to the positions of the sources.

According to Eq.~(\ref{eqn6-2}) and  Eq.~(\ref{eqn6-4}), the complete process of measuring the matrix elements of $\mathsf{D}$ through the fringes and then reconstructing $\mathsf{D}$ by using the duals is described by
\begin{equation}
\label{eqn6-8}
\mathsf{D}'  =   \widetilde{\sf F}^{\rm src}  \mathsf{F}^{{\rm src} \dagger}  \mathsf{D}   \mathsf{F}^{\rm src} \widetilde{\sf{F}}^{{\rm src} \dagger}
\mbox{,}
\end{equation}
where $\mathsf{D}'$ is the reconstruction after the measurement has been made, and we explicitly recognise that $\mathsf{D}$ may not be recovered perfectly because the source fields may not span completely the fields to which the structure is sensitive.

If the measurement set is complete, or overcomplete,  $\mathsf{F}^{\rm src} \widetilde{\sf{F}}^{{\rm src} \dagger}= \mathsf{I}$, and the response $\mathsf{D}$ is recovered perfectly. If the basis is undercomplete, the filter  $\mathsf{F}^{\rm src} \widetilde{\sf{F}}^{{\rm src} \dagger} = {\sf U} {\sf \Sigma} {\sf \Sigma}^{-1} {\sf U}^{\dagger}$ is applied by the measurement and recovery process. Some of the singular values may be too small to be recovered, because of noise, and information will be lost. The operation ${\sf U} {\sf \Sigma} {\sf \Sigma}^{-1} {\sf U}^{\dagger}$ projects the natural modes onto the measurement space, applies a diagonal filter, and then reconstructs the measured modes; information may be lost during this process.

The measurement time increases as the square of the number of individual source positions, but at the outset it is not known how many measurements are needed. There are various ways in which the number of measurements can be minimised. If one source is held at fixed reference position $m$, and the other moved through all possible positions $m'$, the complex visibility observed is given by
\begin{equation}
\label{eqn6-9}
\gamma_{mm'} =  \frac{2 D_{mm'}}{D_{mm} + D_{m'm'}}
\mbox{.}
\end{equation}
It is possible to plot $| \gamma_{mm'} |$ by moving along the $m$'th row of $\mathsf{D}$. Plots such as these reveal the transverse and longitudinal coherence lengths, areas, volumes, and polarisation states of the
collective modes. In an experiment, as one source is held fixed, it is only necessary to scan the second source over the region for which fringes are observed. Another approach is to choose a small, but reasonable, set of sample points, and then to construct the natural modes by calculating the dual matrix and diagonalising the recovered response matrix. New sample points can the be added, and the new dual matrix calculated by using incremental SVD \cite{ref22} on the previous dual matrix. The modes keep on being upgraded as more and more sample points are added until all of the degrees of freedom have been found, and the spatial forms have converged. Similar reasoning can be applied to calculate the cross-response when two different kinds of force are present.

\section{Conclusion}

Energy Absorption Interferometry can be used to gain access to the information contained in the response tensors of many-body systems. It can be implemented at any wavelength; it can be applied to many kinds of generalized force; and it can be used with low-power sources to probe linear behaviour and with high-power sources to probe the differential response of nonlinear systems. It has many advantages compared with attempting to measure the spatial correlations in thermal fluctuations, particularly at low temperatures. In fact, it probes different physics in situations where the fluctuations are not thermal, or where the temperature is not uniform.

In the context of materials characterisation, the complex-valued response tensor can be diagonalised to give the forms and responsivities of the natural dynamical modes through which the material can absorb energy. The technique probes the anti-Hermitian part of the quantum correlation function, and it allows the cross-correlated response to generalized forces of different kinds to be explored. In the case of components such electromagnetic detectors, acoustic sensors, energy harvesting absorbers, and indeed complete instruments, there is no need to deconvolve the illumination patterns of the sources because one is primarily interested in measuring the modal content of the response with respect to some, possibly external, reference surface and source.

The technique can be carried using near-field sources, such as AFM-like (Atomic Force Microscopy) probes, or far-field sources of the kind used in optical and radio antenna test ranges. In the context of studying transport through low-dimensional structures, the close relationship between the Landauer and Kubo formalisms, enables the ports of a sample to be probed by lithographically fabricated leads. Surface acoustic wave transducers oriented at different angles could be used to probe the way in which elastic waves interact with normal metals and superconductors \cite{ref22a}. We are particularly interested in using EAI for probing spin waves, where, unlike FerroMagnetic Resonance, it is not necessary to sweep the field in order to infer modal content from spectra. One can measure modal content at any specific readout frequency and static field strength. An interesting idea is to suspended the sample at the centre of a 3-axis Helmholtz system. A static field can be applied in any direction, and a small superposed modulated field applied in some other direction to interferometrically explore the directional forms of the collective excitations.  There are many different ways in which EAI can be implemented.

A key question is how does one measure the total average power absorbed? In the case of detectors, the output already constitutes the quantity of interest. In the case of general solid-state structures, there are often intrinsic characteristics that are proportional to the power absorbed. For example, in the case of thin-film superconducting resonators, quasiparticle heating leads to a shift in the resonance curve, which is a direct measure of power absorbed \cite{ref23}. Tunnel junctions can also be used to measure the temperatures of electron systems \cite{ref24}. There are numerous other ways of monitoring electron and phonon heating, and these can be implemented depending on the application. We are particularly interested in depositing the material of interest on a suspended dielectric membrane, and then recording the power absorbed by using an ultra-sensitive Transition Edge Sensor \cite{ref25a,ref25b}.

\end{document}